\documentclass[a4paper,11pt]{article}
\usepackage{pos}
\newcommand{\be}{\begin{equation}}
\newcommand{\ee}{\end{equation}}
\newcommand{\bea}{\begin{eqnarray}}
\newcommand{\eea}{\end{eqnarray}}

\title{Axial vector transition form factors in holographic QCD and their contribution to the muon $g-2$}
\ShortTitle{Axial vector TFF in holographic QCD and their contribution to the muon $g-2$}

\author{Josef Leutgeb}
\author*{Anton Rebhan}

\affiliation{Institut f\"ur Theoretische Physik, Technische Universit\"at Wien,\\
        Wiedner Hauptstrasse 8-10, A-1040 Vienna, Austria}

\emailAdd{josef.leutgeb@tuwien.ac.at}
\emailAdd{anton.rebhan@tuwien.ac.at}

\abstract{Whereas the theoretical results for the dominant contributions to hadronic light-by-light scattering 
coming from pseudoscalar meson exchange have converged over the past years, the various published estimates of the contribution due to
axial vector meson exchange differ wildly.
Since holographic AdS/QCD models have proved to provide rather good models of
singly and doubly virtual pion transition form factors, which reproduce remarkably well the known
low-energy data and also the asymptotic leading-order pQCD behavior, it is of interest to consider their predictions for axial vector transition form factors.
Indeed, we find that these reproduce also very well existing data from the L3 experiment
for $f_1\to\gamma\gamma^*$. Including the full infinite tower of axial vector mesons of the AdS/QCD models
we moreover show that the Melnikov-Vainshtein short-distance constraint can be satisfied, while
almost all existing models for hadronic light-by-light scattering fail to incorporate it.
The contribution to $g-2$, which is dominated by the first few resonances, turns out to
be significantly larger than estimated previously.
}

\FullConference{%
  40th International Conference on High Energy physics - ICHEP2020\\
  July 28 - August 6, 2020\\
  Prague, Czech Republic (virtual meeting)
}

\begin{document}
\maketitle

\section{Introduction}

The long-standing discrepancy between the experimental and the theoretical Standard Model (SM) result for the anomalous magnetic moment of the muon $a_\mu\equiv(g_\mu-2)/2$
is currently estimated to be at the level of 3.7 standard deviations
\cite{Aoyama:2020ynm}:
$\Delta a_\mu:=a_\mu^\mathrm{exp}-a_\mu^\mathrm{SM}=279(76)\times10^{-11}$.
The estimated error of the SM result, 
which comes almost exclusively from
hadronic contributions, is dominated by the error in the hadronic vacuum polarization (HVP) part, $40\times 10^{-11}$ according to \cite{Aoyama:2020ynm}. While the hadronic light-by-light (HLBL) scattering contribution is about two orders of magnitude smaller, its uncertainties contribute an error that is about half of the HVP one, $19\times 10^{-11}$. There has been quite some progress in determining the leading HLBL contributions due to the exchange of neutral pseudoscalars with a convergence of refined phenomenological models, the dispersive approach, and lattice calculations (cf.\ the recent review \cite{Aoyama:2020ynm}). However, there is still a great uncertainty concerning the size of the contributions of axial vector meson exchanges and the as we shall see related problem of a failure of most phenomenological models to satisfy the known
short-distance constraints (SDC) of the HLBL tensor, in particular the one implied by the axial anomaly, which has been derived by Melnikov and Vainshtein \cite{Melnikov:2003xd} and which the latter have estimated to require a substantial increase of previous HLBL results.

\begin{figure}[b]
\centerline{\includegraphics[width=.65\textwidth]{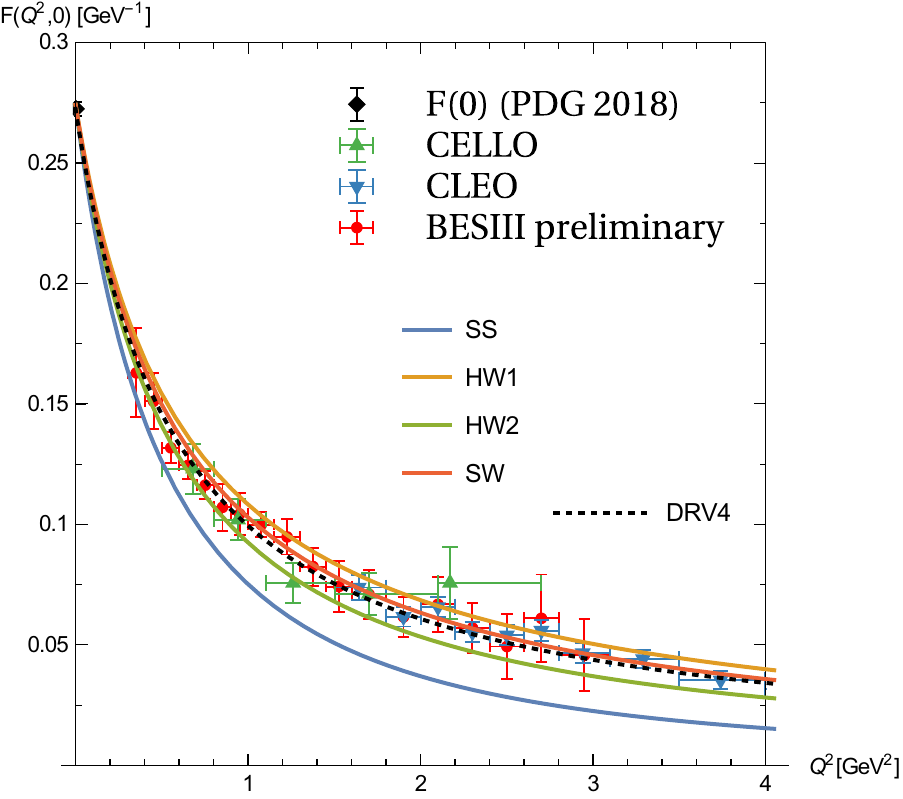}}
\caption{Experimental data for the single-virtual pion TFF compared to the predictions of two hard-wall models HW1, HW2, and a simple soft-wall model (SW) \cite{Leutgeb:2019zpq}. Also included is the prediction of the holographic Sakai-Sugimoto (SS) model, which is a top-down string theoretic construction meant to cover only the low-energy limit of QCD and which misses the short-distance constraints of QCD. The dashed line is a recent fit by Danilkin et al.\ \cite{Danilkin:2019mhd}. (Figure partially taken from Ref.~\cite{Danilkin:2019mhd}.) \label{pionTFF}}
\end{figure}

In \cite{Leutgeb:2019zpq}, we have recently shown that three of the simplest and most popular holographic QCD (hQCD) models,
the two hard-wall models HW1 \cite{Erlich:2005qh,DaRold:2005mxj}, HW2 \cite{Hirn:2005nr}, and a simple soft-wall model (SW) \cite{Karch:2006pv,Grigoryan:2008up,Cappiello:2010uy}, see Fig.~\ref{pionTFF}, reproduce remarkably well the available low-energy data for the single virtual pion transition form factor (TFF) while also reproducing (partially or fully) the short-distance constraints for doubly virtual TFFs due to Brodsky and Lepage (which conventional phenomenological models fail to do, except for the interpolator proposed recently in \cite{Danilkin:2019mhd}). In this paper we report the results obtained in \cite{Leutgeb:2019gbz}, where we obtained axial vector TFFs from these hQCD models and evaluated their contribution to $a_\mu$. 

\section{Axial vector TFFs and $a_\mu$}

\begin{figure}
\includegraphics[width=0.53\textwidth]{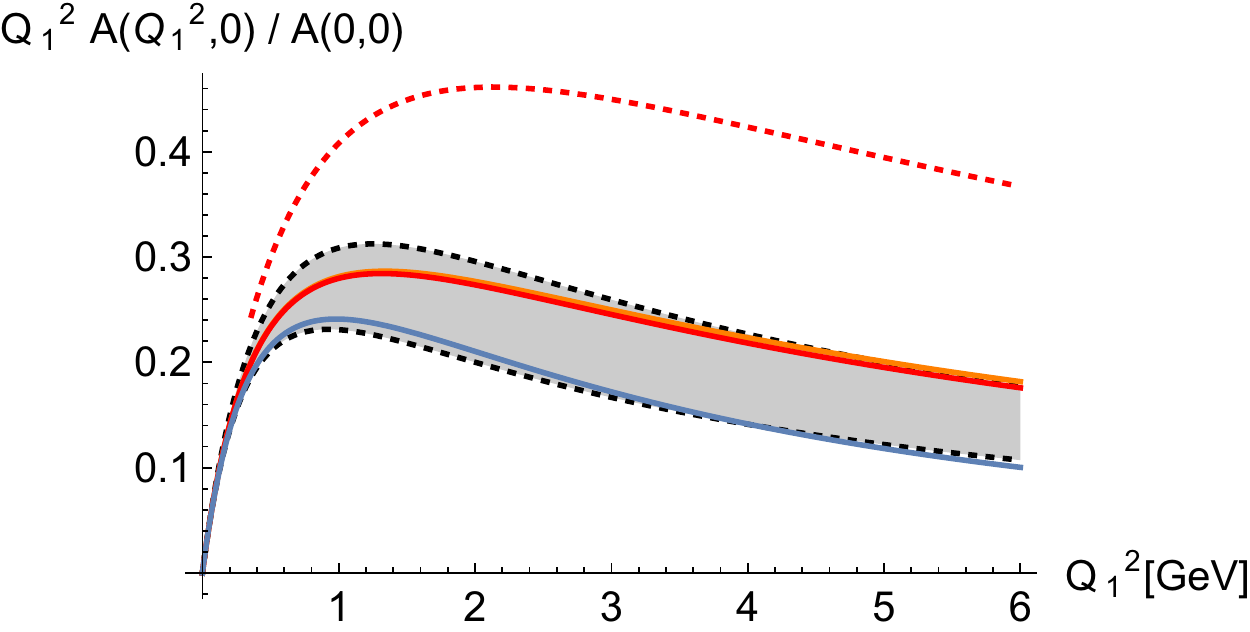}\includegraphics[width=0.47\textwidth]{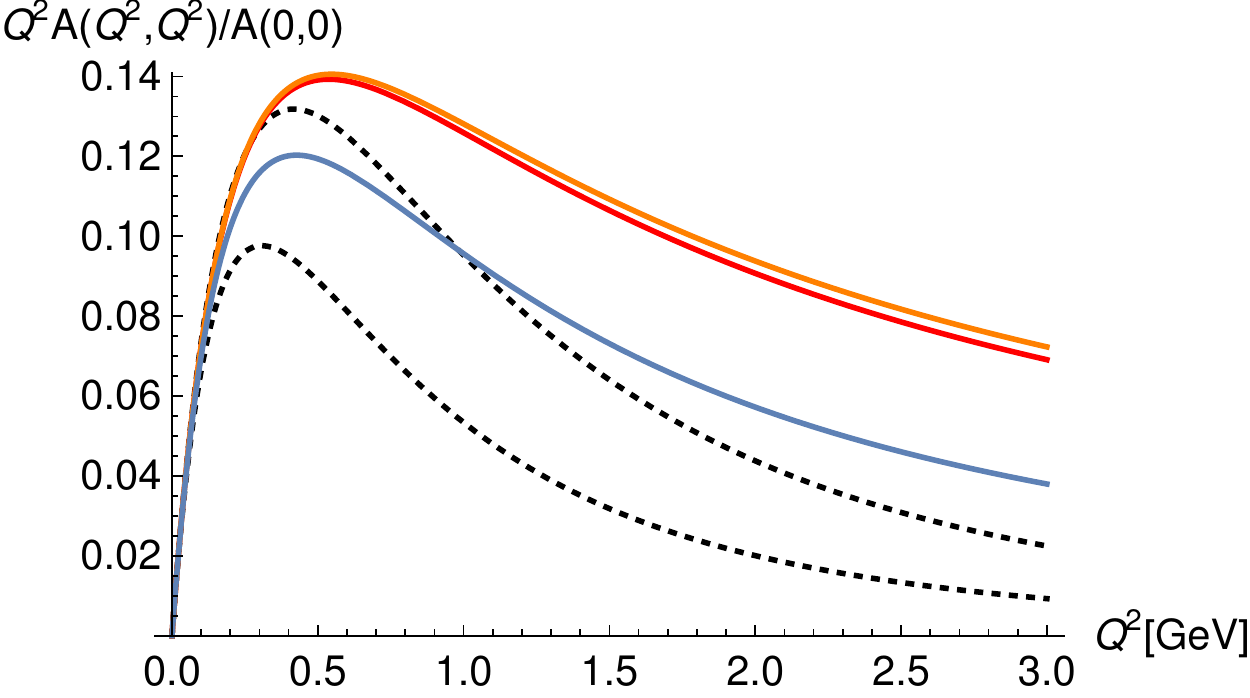}
\caption{Left: Single-virtual axial vector TFFs from holographic models 
(SS: blue, HW1: orange, HW2: red, HW2(UV-fit): red dotted) compared with
dipole fit of L3 data for $f_1(1285)$ (grey band). 
The parameters
of all models except HW2(UV-fit) are fixed by matching $f_\pi$ and $m_\rho$.
(The results for HW1 and HW2 almost coincide here, but $A(0,0)$ differs; see \cite{Leutgeb:2019gbz}.) Right: Double-virtual case with equal virtualities. The black dashed lines correspond to the dipole model used in \cite{Pauk:2014rta}.}
\label{fig:f1}
\end{figure}

In hQCD models, the axial anomaly is included by a five-dimensional Chern-Simons action involving five-dimensional flavor gauge fields whose normalizable modes correspond to the pseudoscalar Goldstone bosons and infinite towers of vector and axial vector fields; photons are included through non-normalizable modes. Pion and axial vector TFFs are therefore unambiguously determined once a hQCD model has been constructed. In hard-wall models where the holographic coordinate $z$ is cut off at $z_0$, the axial vector TFF $A$, corresponding to the
(symmetric) function $A$ in \cite{Pauk:2014rta} has
the form \cite{Leutgeb:2019gbz} (nicely realizing the Lee-Yang theorem by $d\mathcal{J}(0,z)/dz=0$)
\be 
\label{AHW}
A(Q_1^2,Q_2^2) \propto \frac2{Q_1^2} \int_0^{z_0} dz \left[ \frac{d}{dz} \mathcal{J}(Q_1,z) \right]
\mathcal{J}(Q_2,z) \,\psi^A(z),
\ee
where $\mathcal J$ is the bulk-to-boundary propagator of photons and $\psi^A$ is the holographic wave function of one of the axial vector mesons. In contrast to the usual simple model for the axial vector TFF involving a factorized dipole ansatz
\cite{Pauk:2014rta}, the holographic result (\ref{AHW}) is asymmetric in the photon virtualities $Q_1^2$, $Q_2^2$. Its high-energy limit coincides with the asymmetric asymptotic form following from perturbative QCD, which has only most recently been derived in \cite{Hoferichter:2020lap}.

Experimental data for the singly virtual axial vector $f_1(1285)$ TFF have been obtained by the L3 experiment \cite{Achard:2001uu} in the form of a dipole fit, which is represented by the grey band in the left panel of Fig.~\ref{fig:f1} and compared with the results from the HW1 and HW2 models as well as the top-down SS model. The latter are right on top of the experimental results within errors. As shown in right panel of Fig.~\ref{fig:f1}, for the doubly virtual case (where no experimental data are so far available)
the holographic prediction deviates substantially from the simple ansatz used in \cite{Pauk:2014rta}.
As a result, the holographic result for the axial-vector $f_1(1285)$ contribution to $a_\mu$ deviates significantly from \cite{Pauk:2014rta}.

\begin{figure}
\hspace*{-1.25cm}\includegraphics[width=0.555\textwidth]{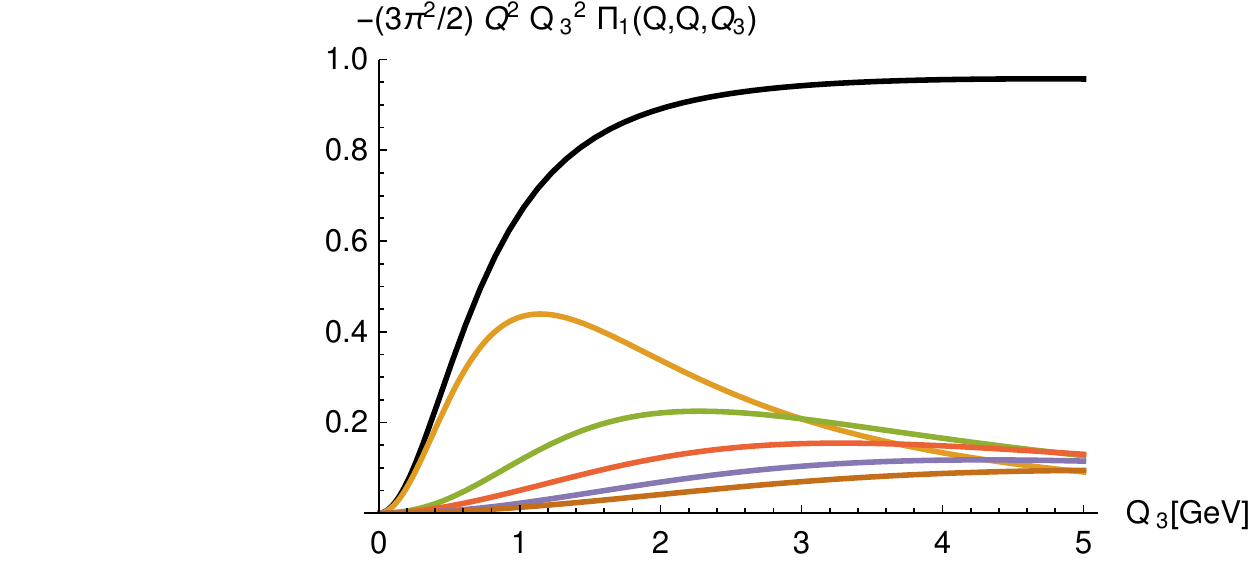}
\includegraphics[width=0.5\textwidth]{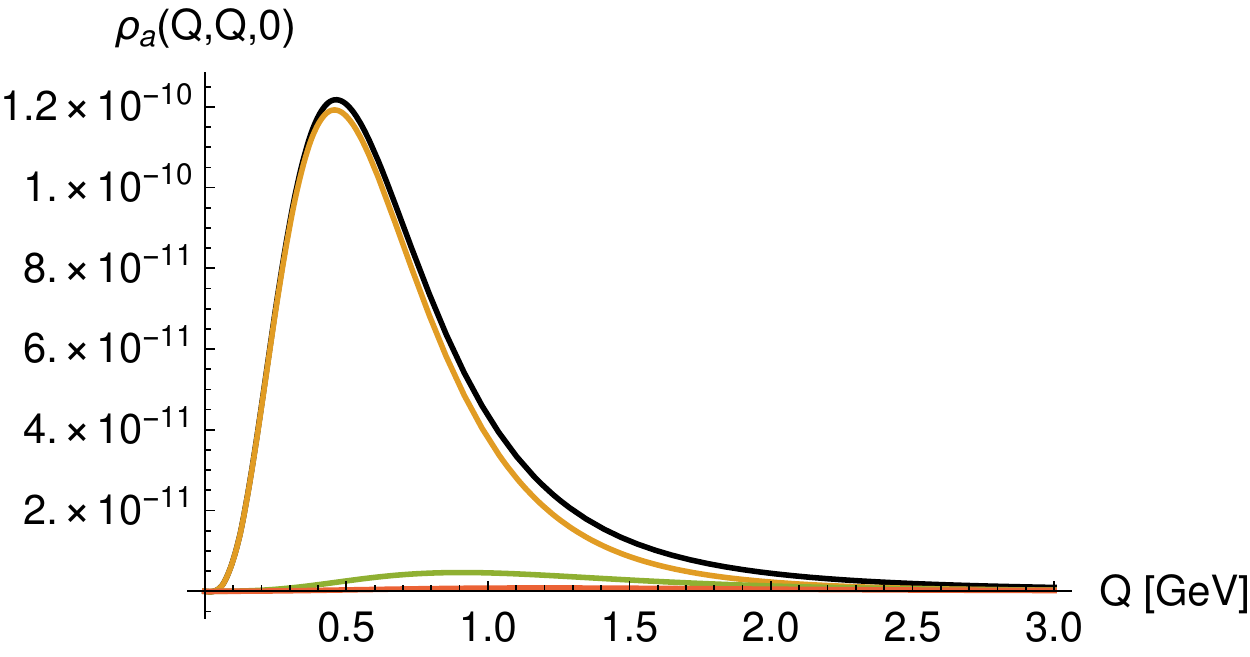}
\caption{Axial-vector contribution to
$Q_3^2 Q^2 \bar\Pi_1(Q,Q,Q_3)$ as a function of $Q_3$ at $Q=50$ GeV
in the HW2 model and the integrand of $a_\mu^\mathrm{AV}$ \cite{Leutgeb:2019gbz}. 
The black line corresponds to the infinite sum over the tower of axial vector mesons, and the other lines
give the contributions of the 1st to 5th lightest axial vector mesons.
}
\label{fig:MV}
\end{figure}

A theoretically even more important aspect of the holographic result for the axial-vector contributions to $a_\mu$ 
arises when the whole tower of axial vector mesons present in hQCD models is summed up. In the left panel of Fig.~\ref{fig:MV} the component of the HLBL tensor involved in the Melnikov-Vainshtein SDC is considered, showing that each individual axial vector mode gives a vanishing contribution at infinite momenta, but the infinite sum does approach a finite value as required by the SDC.
However, as shown in the right panel of Fig.~\ref{fig:MV}, the integrand $\rho_a$ of the two-loop integral for $a_\mu$ receives significant contributions only from the first few axial vector modes. The corresponding numerical results for $a_\mu$ are given in Table \ref{tab:amuj}.

\begin{table}
\centering
\begin{tabular}{l|c|c|c|c|c|c|}
 & $j=1$ & $j\le2$ & $j\le3$ & $j\le4$ & $j\le5$ & $a_\mu^\mathrm{AV}$ \\
\hline
HW1 & 31.4 & 36.2 & 37.9 & 39.1 & 39.6 & \phantom{\Big|}40.6 $\times 10^{-11}$ \\
HW2 & 23.0 & 26.2 & 27.4 & 27.9 & 28.2 & \phantom{|}28.7 $\times 10^{-11}$ \\
\hline
\end{tabular}
\caption{The contribution of the infinite tower of axial vector mesons to $a_\mu^\mathrm{AV}$ in the two hard-wall AdS/QCD models considered in \cite{Leutgeb:2019gbz}.)
}
\label{tab:amuj}
\end{table}

\section{Discussion}

In Table \ref{tab:amutotal}
the results for $a_\mu^\mathrm{PS}$ obtained in \cite{Leutgeb:2019zpq} and the new results for $a_\mu^\mathrm{AV}$ are combined for the two hard-wall models considered by us. The parameters of these models have been fixed to reproduce the right values for the $\rho$ meson mass and the pion decay constant which led to the results shown in Fig.~\ref{pionTFF}. Doing so, the model HW2 cannot be matched to the full value of the various asymptotic constraints, but instead satisfies them at the level of only 62\%.
Model HW1 contains one more free parameter so that the SDCs can be satisfied fully. However, neither has a running coupling constant. Therefore at large but finite momenta where NLO effects are important, the actual behavior in real QCD 
may well be in between that of HW1 and HW2, so that the two holographic results may be taken to delimit a plausible range of predictions.\footnote{In Ref.~\cite{Cappiello:2019hwh} results similar to ours have been obtained by using only the HW2 model,
where two sets of parameters have been used that also span the range of 62\% and 100\% SDC saturation. For the latter the mass of the rho meson can no longer be fitted, so that neither the pion TFF nor the axial vector TFF is reproduced well (the latter is displayed by the red dotted line in Fig.~\ref{fig:f1}). In fact, both choices yield a result for $a_\mu^\mathrm{AV}$ which agrees with our HW2 result.}

\begin{table}
\centering
\begin{tabular}{l|c|c|}
 & HW1 (100\% MV-SDC) & HW2 (62\% MV-SDC) 
 \\
\hline
$a_\mu^\mathrm{PS}[\pi^0+\eta+\eta']\times 10^{11}$ & 92.2 [61.3+16.7+14.2] & 83.7 [59.2+15.9+13.4] \phantom{\Big|} 
\\
$a_\mu^\mathrm{AV}[L+T]\times 10^{11}$ & 40.6 [23.2+17.4] & 28.7 [16.6+12.0] \phantom{\Big|} 
\\
\hline
$a_\mu^\mathrm{PS+AV}\times 10^{11}$ & 133 & 112 \phantom{\Big|} 
\\
\hline
\end{tabular}
\caption{Summary of the results of our previous calculation of the pseudoscalar pole contribution of Ref.~\cite{Leutgeb:2019zpq}
and the results obtained in \cite{Leutgeb:2019gbz} for the contribution of axial vector mesons in the models HW1 and HW2.
}
\label{tab:amutotal}
\end{table}

In the White Paper \cite{Aoyama:2020ynm} the contribution from axial vector mesons and the SDC have been estimated as
$a_\mu^\mathrm{SDC}=15(10)\times 10^{-11},
$
$
a_\mu^\mathrm{axials}=6(6)\times 10^{-11},$
and it was suggested to add their errors linearly. The sum,
$21(16)\times 10^{-11}$, may be compared to our combined HW1-HW2 results \cite{Leutgeb:2019gbz}\footnote{In Ref.~\cite{Leutgeb:2019gbz} we have also considered a phenomenological matching of the holographic axial vector TFF to the L3 results resulting in $a_\mu^\mathrm{AV}=22(5)\times 10^{-11}$ at the expense of the SDC . The latter deficiency should then be compensated by excited pseudoscalar contributions (which are not present in our HW models, because they are chiral), so that a final value close to (\ref{aAVrange}) may result. In fact, in Ref.~\cite{Colangelo:2019uex} a model involving an infinite tower of pseudoscalars has been constructed to account for the Melnikov-Vainshtein SDC.}
\be\label{aAVrange}
a_\mu^\mathrm{AV}[L+T]=35(6)\,[20(3)+15(3)]\times 10^{-11},
\ee
where the longitudinal (L) part
is responsible for the saturation of the Melnikov-Vainshtein SDC.
The holographic results thus overlap those of the White Paper \cite{Aoyama:2020ynm} within errors, but are significantly\footnote{An increase of $a_\mu^\textrm{axials+SDC}\times10^{11}$ from 21 to 35 would reduce the current discrepancy between theory and experiment from 3.7 to 3.5$\sigma$.} larger, in particular as concerns the contribution from (transverse) axial vector mesons.\footnote{Note, however, that estimates such as the one of Ref.~\cite{Pauk:2014rta} did not remove the longitudinal axial-vector contributions.}
In our opinion, the hQCD models provide a more plausible
model for axial vector TFFs than previous phenomenological models,
because they reproduce rather well the experimental results at
low energies while matching
the very nontrivial doubly virtual form \cite{Hoferichter:2020lap} of leading-order perturbative QCD at high energies.
It would be very interested to have them further tested by experiment.

\begin{acknowledgments}
J.~L.\ was supported by the FWF doctoral program
Particles \& Interactions, project no. W1252.
\end{acknowledgments}

%

\raggedright
\bibliographystyle{../JHEP}
\bibliography{../hlbl}

\end{document}